\documentclass[10pt]{iopart}

\usepackage{cite}
\usepackage{iopams}
\usepackage{graphicx}
\usepackage{bbm}

\newcommand{\R}{\mathbbm{R}}
\newcommand{\D}{\Delta}
\renewcommand{\L}{\Lambda}
\newcommand{\eg}{\textit{e.g.}\ }
\newcommand{\ie}{\textit{i.e.}\ }

\newtheorem{theorem}{Theorem}
\newtheorem{lemma}[theorem]{Lemma}

\begin{document}

{\hspace*{\fill} Preprint-KUL-TF-2002/05}

\letter{Proof of Bose-Einstein Condensation for Interacting Gases with a
One-Particle Spectral Gap} 

\author{J. Lauwers, A. Verbeure, V. A. Zagrebnov\footnote[7]{on leave of absence from 
Universit\'e de la M\'editerran\'ee and Centre de Physique Th\'eorique, 
CNRS-Luminy-Case 907, 13288 Marseille, Cedex 09, France
}}

\address{Instituut voor Theoretische Fysica,  
Katholieke Universiteit Leuven,   
Celestijnenlaan 200D,   
B-3001 Leuven, Belgium}

\eads{
\mailto{joris.lauwers@fys.kuleuven.ac.be},
\mailto{andre.verbeure@fys.kuleuven.ac.be},
\mailto{zagrebnov@cpt.univ-mrs.fr}
}

\begin{abstract}
Using a specially tuned  mean-field Bose gas as a reference system, we
establish a positive lower bound on the condensate density for continuous Bose
systems with superstable two-body interactions and a finite gap in the one-particle
excitations spectrum, \ie we prove for the first time standard homogeneous
Bose-Einstein condensation for such interacting systems. 
\end{abstract}


%

\section{Introduction}

The long-standing problem of proving the existence of a Bose-Einstein
condensation (BEC) in non-ideal real Bose gases with a standard two-body 
interaction has recently been warmed up by the great success of observing this 
phenomenon in trapped gases. Notice that much earlier BEC has been observed in 
liquid $\mathrm{{}^4He}$, which however remained always under discussion.

In the present note we announce our result together with a sketch of the proof
about the existence of the standard or zero-mode BEC 
in Bose gases with realistic superstable 
two-body interactions and with gap in the one-particle excitation energy 
spectrum. To the best of our knowledge, this is a first proof of its sort for 
homogeneous systems. We are not using any scaling limits, (\eg type van der 
Waals limits \cite{lebowitz:1966,lieb:1966,buffet:1983b}) or truncation of 
particle interactions \cite{dorlas:1993,zagrebnov:2001}. We prove that BEC 
occurs by constructing a positive lower bound for the condensate density which 
is valid for low enough temperature and appropriate large density of particles.
 
We consider a gas of interacting Bosons in cubic boxes $\L = L
\times L \times L \subset \R^3$ with periodic boundary conditions. 
We look here in detail at the $3$ dimensional case, and comment later
on the case of other dimensions. Denote by $V= L^3$ the
volume of the box $\L$. The grand-canonical Hamiltonian of this system reads
\begin{equation} \label{H-int}
H^\D_{\L, g} = T^\D_\L -\mu N_\L + g U_\L, \qquad g > 0,
\end{equation} 
where $T^\D_\L$ is the kinetic energy with gap $\Delta > 0$ in its spectrum,
\begin{equation}\label{T-Delta}
T^\D_\L = \sum_{k \in \L^*} \frac{\hbar^2k^2}{2m} a^\dagger_k a_k -\Delta
a^\dagger_0 a_0.
\end{equation}
The sum $k$ runs over the set $\L^*$, dual to $\L$, \ie 
\begin{equation*}
\L^* = \left\{ k
\in \R^3; k_\alpha = 2\pi n_\alpha /L; n_\alpha = 0,\pm 1,\ldots; \alpha
=1,2,3 \right \}.
\end{equation*} 
 The operators $a^\dagger_k$ and $a_k$ are the Bose
creation and annihilation operators for mode $k$. As usual, the mode-occupation 
number operators are denoted by $N_k = a^\dagger_ka_k $, 
$N_\L = \sum_{k \in \L^*}N_k$ is the total number operator in $\L$.

We assume \textit{a priori} the presence of a gap $\D$ in the one-particle
excitations spectrum, isolating the lowest (zero-mode) energy level. 
This can be realised by taking appropriate boundary conditions, 
attractive boundary conditions \cite{robinson:1976,landau:1979}, or such a gap 
can also be realised by specific interparticle interactions effectively 
incorporated in general two-body interactions \cite{zagrebnov:2001}.

Of course $\mu$ is the chemical potential and
the interaction between the particles is modelled by the two-body interaction
term
\begin{equation}\label{U-Lambda}
U_\Lambda =  \frac{1}{2}\int_{\Lambda^2}\!\mathrm{d}x\mathrm{d}y\ a^\dagger (x)
a^\dagger (y) v( x - y) a(y)a(x),
\end{equation}
where 
$a^\dagger (x),a^\dagger (y)$ and $a(y),a(x)$ are the creation and annihilation 
operators for the Bose particles at $x,y \in \R^3$. The interaction potential 
$v$ is assumed to be 
spherically symmetric,  
superstable \cite{ruelle:1969}, \ie it satisfies the inequality
\begin{equation}\label{s-stable}
\sum_{1 \leq i < j \leq n}v(x_i - x_j) \geq 
\frac{A}{2V}n^2 - B n
\end{equation}
for some constants $A > 0$, $B \geq 0$, and all $n \geq 2$, $x_i \in \L$.
Consequently, the interaction term (\ref{U-Lambda}) satisfies
\begin{equation}\label{sstab2}
U_\L \geq \frac{A}{2V}N_\L^2 - B N_\Lambda. 
\end{equation}
This superstability property is, together with the spectral gap (\ref{T-Delta}), 
the physical foundation of our proof.  Intuitively, one might understand that 
condensation in the groundstate (\ie $k=0$), which is energetically isolated by 
a gap $\Delta$ can survive the switching-on of a gentle interaction, 
and that fluctuations must be of a macroscopical size to overcome this gap 
and lift particles out of the isolated groundstate.
More technical ingredients of the proof are the convexity
properties of thermodynamical potentials, such as the pressure, and the
use of an optimal choice for the constants $A,B$ in the superstability
criterion (\ref{s-stable}). Indeed, it was proved \cite{ruelle:1969} that 
continuous $L^1$-functions of positive type $v : \R^3 \to \R$ are superstable 
potentials if and only if
\begin{equation}\label{vhat-0}
 \hat{v}(0) \geq \hat{v}(q) = \int_{\R^3}\! \mathrm{d}x\ v(x) \mathrm{e}^{-
 \mathrm{i} q x}
 \geq 0,\qquad \forall q \in \R^3,
\end{equation}
and $\hat{v}(0)> 0$.
Moreover, Lewis, Pul\`e, and de Smedt \cite{lewis:1984} proved the existence of 
the optimal constants $A = \hat{v}(0)(1 - \epsilon)$ and $B = v(0)/2$ in 
(\ref{s-stable}) for this type of potentials. Here, $\epsilon > 0$ is an arbitrarily
small positive constant. 
It is related to the size of the system, and can be put to zero after the
thermodynamic limit. This optimal choice is of determining 
importance in our proof of the zero-mode Bose-Einstein condensation.

\section{Sketch of Proof}

The main idea of the proof is to estimate the Bose condensate of the full
model (\ref{H-int}) by the condensate of a particularly chosen reference system
for which one knows the occurrence of condensation. The clever choice of this 
reference system is the subtle point of our proof. 
The reference system is a so-called mean-field Bose gas, an exactly solvable 
model of Bosons \cite{huang:1967, davies:1972, fannes:1980b, buffet:1983,
berg:1984,papoyan:1986, lewis:1988} (for a review see \cite{zagrebnov:2001}) 
defined by the grand-canonical Hamiltonian as
\begin{equation}\label{H-MF}
H^{\Delta}_{\Lambda,g,\lambda} = T_\Lambda^\Delta -\mu N_\Lambda +
g\frac{\lambda}{2V}N_\Lambda^2.
\end{equation}
The kinetic energy operator $T_\Lambda^\Delta$  (\ref{T-Delta}) is as 
for the general interacting system (\ref{H-int}), but the interaction term
(\ref{U-Lambda}) is replaced by a mean-field interaction term.

The reference system (\ref{H-MF}), a mean-field Bose gas, emerges as 
the \textit{van der Waals} limit of the fully interacting system
\cite{lebowitz:1966,lieb:1966, buffet:1983b}.  In that case,
the constant $\lambda$ equals $\hat{v}(0)$ (\ref{vhat-0}), which means that the
van der Waals limit reduces the full interaction to the $\hat{v}(0)$
contribution. In our proof we tune the constant $\lambda$ in order to get the
best possible lower bound for the condensate density of the full interaction
model.

Remark that our reference system (\ref{H-MF}) does show Bose condensation for
large enough densities (\ie for $\mu$ large) at any given temperature. Moreover,
systems of the type of our reference system have better properties than the
ideal Bose gas which is in many ways a \textit{pathological} model, \eg in the
sense that there is no equivalence of ensembles \cite{cannon:1973,lewis:1974} 
and in the sense that the chemical potential is limited by a zero upperbound in order to 
safeguard thermodynamical stability.  These are the reasons for our strategy of 
using the free of those pathologies reference system (\ref{H-MF}).

\textit{Thermodynamic properties of the reference system (\ref{H-MF}).} It is
well known that our reference system (\ref{H-MF}) is a soluble model. In the
case of vanishing gap $\D = 0$, the complete solution can be found at several
places in the literature \cite{huang:1967,davies:1972,fannes:1980b,buffet:1983,
berg:1984,papoyan:1986,lewis:1988,zagrebnov:2001}. In particular there is
condensation for all dimensions $D \geq 3$ at any temperature  for densities
large enough. It is a student exercise to work out now the case with gap 
$\D > 0$. There is one main difference with the gapless case, namely the
presence of the gap provokes a shift in the chemical potential and its
treshholds, and one gets condensation in all dimensions $D \geq 1$. One derives
straightforwardly that for the reference model (\ref{H-MF}) one gets
condensation for all values of the chemical potential satisfying,    
\begin{equation*}
\mu > g\lambda \rho^{P}(\beta,-\D) -\D.
\end{equation*}
where $\rho^{P}(\beta,-\D)$ is the total particle density of the perfect Bose Gas
(PBG) at inverse temperature $\beta$ and chemical potential equal to $-\D$. 
Moreover, the condensate density  
\begin{equation*}
\rho^{\D }_{0,g, \lambda}(\beta,\mu) = \lim_{V \to \infty}\frac{1}{V}
 \langle N_0 
\rangle_{H_{\L,g, \lambda}^\D}(\beta,\mu),
\end{equation*}
\ie the particle density at the zero mode in the thermodynamic limit 
$(V \to \infty)$ of 
the grand-canonical Gibbs states $\langle - \rangle_{H_{\L}}{(\beta,\mu)}$, in 
volumes $\L$ for a certain choice of inverse temperature and chemical 
potential $(\beta,\mu)$ and Hamiltonian $H_\L$, is
explicitly given by
\begin{equation}\label{r0-mf-D} 
  \rho^{\D}_{0,g,\lambda} = \frac{\mu + \D}{g\lambda}
  -\rho^{P}(\beta,-\D).
\end{equation}  
One also computes the total particle density for given $(\beta,\mu)$ to be
\begin{equation}\label{r-mf-D}
  \rho^{\D}_{g,\lambda}  =  \lim_{V \to \infty}\frac{1}{V}\langle 
 N_\L \rangle_{H_{\L,g,\lambda}^{\D}} = \frac{\mu + \D}{g\lambda}.
\end{equation}

To prove Bose condensation in the full model (\ref{H-int}), we subtract from
(\ref{U-Lambda}) the long-range part of the interaction proportional to 
$N_\L^2/2V$, tune it with a factor, taking into account the optimal stability 
constants, and add it to the kinetic-energy term. The latter serves as our reference 
system (\ref{H-MF}), from which we establish a lower bound on the zero-mode
condensate density $\rho^\Delta_{0,g}(\beta,\mu) =\lim_{V \to \infty}\langle 
N_0/V\rangle_{H_{\Lambda, g}^{\D}}$ in the fully interacting system (\ref{H-int}).  
In the lemma below, a lower bound on $\rho^\Delta_{0,g}(\beta,\mu)$ is given. 
\begin{lemma}\label{lemma-lb}
The zero-mode condensate density $\rho^\Delta_{0,g}(\beta,\mu)$ in the thermodynamic limit
of grand-canonical Gibbs states of interacting systems (\ref{H-int}) with 
superstable two-body potentials $v$ (\ref{vhat-0}), has the following lower
bound:
\begin{eqnarray}\nonumber
\rho^\Delta_{0,g}(\beta,\mu) &\geq&  \frac{\mu}{g\hat{v}(0)} +
\frac{g\hat{v}(0)}{2\Delta}\left(\rho^{P}(\beta,-\Delta)\right)^2 
- \frac{\mu +\Delta}{\Delta}\rho^{P}(\beta,-\Delta)  
\\ && -\ \frac{g v(0)}{2\Delta}\rho^{(\Delta = 0)}_g(\beta,\mu) - \rho^{P}_c(\beta).
\label{lb}
\end{eqnarray}
Here $\rho_g^{(\Delta = 0)}(\beta,\mu)$ is the total
density of the interacting gas without gap (\ref{H-int}). 
$\rho^{P}(\beta,-\Delta)$ refers to the total 
density of the PBG at inverse temperature $\beta$ and chemical potential 
$\mu = -\D$, $\rho_c^{P}(\beta)$ is the critical density of the PBG. The bound 
is valid for values $\mu > g\hat{v}(0)\rho_c^{P}(\beta)$, and
dimensions $D \geq 3$.
\end{lemma}

\textit{Idea of the proof}.
Using the Bogoliubov convexity inequality \cite{zagrebnov:2001} one gets:
\begin{equation}\label{b-conv}
\frac{g}{V}\langle W_\L^\lambda \rangle_{H^{\D}_{\L,g}} \leq
p_\Lambda[H^{\D}_{\L,g,\lambda}] - p_\Lambda[H^{\Delta}_{\Lambda,g}]
\leq \frac{g}{V}\langle W_\Lambda^\lambda \rangle_{H^{\D}_{\L,g,\lambda}}.
\end{equation}
This gives upper and lower bounds on the difference of the grand-canonical 
pressure $p_\L[H_\L]$ of the mean-field reference Bose gas (\ref{H-MF}) and 
the full model (\ref{H-int}).  The operator $W_\L^\lambda$ is the difference 
between the interactions of the fully interacting and the mean-field Bose gases,
\ie $W_\Lambda^\lambda = U_\Lambda - \frac{\lambda}{2V}N_\L^2$.   
The expectation values in (\ref{b-conv}) can be estimated using, 
for the lower bound, the superstability properties of the interaction, and, 
for the upper bound, the properties of the equilibrium states of the 
mean-field reference Bose gas.
The lower bound in (\ref{b-conv}) follows from  (\ref{sstab2}), and from
the tuning of the interaction parameter $\lambda$ for the mean-field reference 
Bose gas (\ref{H-MF}) to the constant $A$ in (\ref{sstab2}),
\begin{equation}\label{b1}
\frac{g}{V}\langle W_\Lambda^{A} \rangle_{H^{\D}_{\L,g}} \geq 
-\frac{gB}{V}\langle N_\Lambda \rangle_{H^{\D}_{\L,g}}.
\end{equation}  
On the other hand, using the mode by mode gauge invariance of the Gibbs states of the mean-field
Bose gas we arrive at the following upper bound for the pressure difference
(\ref{b-conv}),
\begin{equation}\label{b2}
\frac{g}{V}\langle W_\L^{A} \rangle_{H^{\D}_{\L,g,A}}
\leq  \frac{g}{V^2} \langle C N_\L^2 
- \frac{\hat{v}(0)}{2}N_0^2 \rangle_{H^{\D}_{\L,g,A}},
\end{equation}
where $C = \hat{v}(0) -A/2$.
It follows from the properties of the mean-field Bose gas that the expectation
values in the rhs of (\ref{b2}) in the limit $(V\to\infty)$ are
given by $gC(\rho^{\D}_{g,A}(\beta,\mu))^2 - 
g\hat{v}(0)(\rho^{\D}_{0,g,A}(\beta,\mu))^2/2$.

The pressure $p_\L[H^\D_\L]$ is an increasing convex function of
$\D \geq 0$. Since the condensate density $\rho^\Delta_{0,g}(\beta,\mu)$ is 
the derivative 
of the pressure with respect to $\D$, by convexity we 
find  for it a lower bound, given by
\begin{equation}\label{c1}
\frac{1}{V}\langle N_0 \rangle_{H_{\L, g}^{\D}} 
\geq  \frac{p_\L[H^\D_{\L,g}] - 
p_\L[H^{(\D =0)}_{\L,g}]}{\D}.
\end{equation}
Analogously, by virtue of the same convexity property, 
the difference of the pressures between the mean-field Bose gas 
with gap and without gap, is bounded from below by the condensate density 
for the mean-field gas without gap, \ie
\begin{equation}\label{c2}
\frac{p_\Lambda[H^{\D}_{\L,g,A}] - p_\L[H^{(\D =0)}_{\L,g,A}]}{\D} \geq 
\frac{1}{V}\langle N_0 \rangle_{H_{\L,g,A}^{(\D=0)}} .
\end{equation}
Adding up these two inequalities and using the Bogoliubov convexity inequality
(\ref{b-conv}) once at $\D >0$ and once at $\D = 0$, together with the
bounds (\ref{b1}) and (\ref{b2}), one gets the 
following lower bound for the condensate density $\rho^\D_{0,g}(\beta,\mu)$ 
in the thermodynamic limit ($V \to \infty$):
\begin{eqnarray}\nonumber
\rho^\D_{0,g}(\beta,\mu)  &\geq&  \rho^{(\D=0)}_{0,g,A}(\beta,\mu)
+  g\frac{\hat{v}(0)}{2\Delta}(\rho^{\Delta}_{0,g,A}(\beta,\mu))^2
\\& & -\ \frac{g}{\D}\left( B \rho^{(\Delta=0)}_{g}(\beta,\mu) +
C(\rho^{\D}_{g,A}(\beta,\mu))^2 \right).\label{lb-1}
\end{eqnarray}
The lower bound (\ref{lb}) now follows from (\ref{lb-1}) by use
of the explicit expressions for total density and the condensate 
density of the mean-field Bose gas with gap, (\ref{r-mf-D}) and (\ref{r0-mf-D}),
and by the well-known expression for the condensate density in the gapless
mean-field model:  
\begin{equation*}
\rho^{(\D =0)}_{0,g,\lambda} = \lim_{V \to \infty}\frac{1}{V}\langle N_0
  \rangle_{H_{\L,g,\lambda}^{(\D =0)}} = \frac{\mu}{g\lambda}
  - \rho_c^{P}(\beta),
\end{equation*}
if $\mu > g \lambda\rho^{P}_c(\beta)$,
where $\rho^{P}_c(\beta)$ is the critical density for the PBG at
inverse temperature $\beta$.
As a last step, we need the optimal superstability constants for continuous 
$L^1$ potentials of positive type \cite{lewis:1984}, \ie we take
$A = \hat{v}(0)$, and $B = v(0)/2$ after the thermodynamic limit, 
and get the expression for the lower bound (\ref{lb}) in the lemma.
\hfill \textit{QED}

Now we get our main result:
\begin{theorem}\label{theorem}
Consider a three dimensional system of interacting Bose particles (\ref{H-int}), with
a superstable two-body potential.
There exists  a minimal gap $\Delta_{min}$ such that for any finite 
$\Delta \geq \Delta_{min}$, one has $\rho^\Delta_{0,g}(\beta,\mu) > 0$ or
zero-mode condensation.
\end{theorem}
\textit{Proof.} 
Take any $\eta > 0$, fix a temperature and a chemical potential $(\beta,\mu)$
such that $\mu > g\hat{v}(0)(\rho^{P}_c(\beta) + 3 \eta)$,
then there exists  a minimal gap $\Delta_{min}$ such that for any finite 
$\Delta \geq \Delta_{min}$,
\begin{equation*}
\left|\frac{g\hat{v}(0)}{2\Delta}\left(\rho^{P}(\beta,-\Delta)\right)^2  
- \frac{g v(0)}{2\Delta}\rho_g^{(\Delta = 0)}(\beta,\mu) 
- 2\frac{\mu +\Delta}{\Delta}\rho^{P}(\beta,-\Delta)\right| < \eta
\end{equation*} 
For these values of the gap $\Delta$ we have 
$\rho^\Delta_{0,g}(\beta,\mu) > \eta > 0$, by virtue of the lower bound
(\ref{lb}) in Lemma~\ref{lemma-lb}, and hence we proved condensation.
\hfill \textit{QED}

\section{Discussion}
First of all let us remark that our proofs hold 
without any change in all dimensions $D \geq 3$. For $D = 1$ or $2$ a similar 
lower bound on the condensate density can be derived, on the basis of modified 
convexity arguments (\ref{c1})--(\ref{c2}), \ie one has to consider pressure 
differences of the form $p_\Lambda[H^{\D}_{\L}] - p_\L[H^{\D_0}_{\L}]$, with 
$0 < \D_0 < \D$, instead of with $\D_0 =0$ in (\ref{c1})--(\ref{c2}). This yields 
the substitution of $\rho^{\D_0}_{0,g,A}(\beta,\mu)$ and
$\rho_g^{\D_0}(\beta,\mu)$ for $\rho^{(\D=0)}_{0,g,A}(\beta,\mu)$ and
$\rho_g^{(\D=0)}(\beta,\mu)$ in (\ref{lb-1}). Hence, also in one and two 
dimensional interacting Bose gases with gap (\ref{H-int}), Bose 
condensation is proved, in contrast to the Bogoliubov--Hohenberg 
theorem \cite{bratteli:1996} which yields the absence of BEC for 
$1D$ or $2D$ translation invariant continuous Bose systems without gap.   

Finally we want to stress that our results are for homogeneous systems. On the
other hand we have to mention here the interesting recent result about the
condensation for trapped Bose gases \cite{lieb:2002}, \ie for inhomogeneous
systems where a rigorous proof is given of Bose condensation in the
so-called Gross-Pitaevskii limit. As such trapped systems always have a gap in
the one-particle spectrum, we consider our result as a bridge between the
homogeneous gapless systems on one hand and the trapped systems on the
other hand.  Moreover, we hope that our work might be inspiring to establish  
a proof of BEC for homogeneous systems when the gap tends to zero.

\ack{ J. L. gratefully acknowledges financial support from K.U.Leuven 
 grant FLOF-10408, and V. A. Z. acknowledges ITF K.U.Leuven for hospitality.
}
\\

\end{document}